# A medium-resolution monochromator for 73 keV x-rays - Nuclear resonant scattering of synchrotron radiation from $^{193}$Ir

**Pavel Alexeev[a]\*, Frank-Uwe Dill[a], Hans-Christian Wille[a]\*, Ilya Sergeev[a], Marcus Herlitschke[a], Olaf Leupold[a], Desmond F. McMorrow[b] and Ralf Röhlsberger[a]**

[a]Photon Science, Deutsches Elektronen-Synchrotron DESY, Notkestr. 85, Hamburg, Hamburg, 22607, Germany

[b]London Centre for Nanoscience and Department of Physics and Astronomy, University College London, WC1E 6BT, London, United Kingdom

Correspondence email: pavel.alexeev@desy.de; hans.christian.wille@desy.de

**Abstract** We report on the development and characterization of a medium resolution monochromator for synchrotron-based hyperfine spectroscopy at the 73 keV nuclear resonance of $^{193}$Ir. The device provides high throughput of $6*10^8$ ph/s in an energy bandwidth of 300(20) meV. We excited the nuclear resonance in $^{193}$Ir at 73.04 keV and observed nuclear fluorescence of $^{193}$Ir in Iridium metal. The monochromator allows for Nuclear Forward Scattering spectroscopy on Ir and its compounds.

**Keywords:** X-ray optics, Nuclear resonance scattering, Iridium, APD detectors

## Introduction

The usefulness of modern accelerator based x-ray sources directly depends on the quality and the performance of the implemented x-ray optics. For highly efficient experiments the latter have to provide collimation, focusing and monochromatization while preserving the spectral flux delivered by the source. High energy resolution in combination with high throughput is of largest importance for flux-hungry techniques which are those that exhibit small scattering cross-sections, such as magnetic scattering or inelastic x-ray scattering, or resonant techniques with very narrow spectral linewidths like nuclear resonant scattering (NRS) [1,2]. Nuclear forward scattering (NFS), the Mössbauer effect based hyperfine spectroscopy in the time domain, has a very broad range of applications. For example, it is used to study magnetic moments in nano-materials, determine the oxidation state during chemical processes or to study quantum optical phenomena in the X-ray regime [2-5].

The main challenge in a NRS experiment is to detect resonantly scattered single photons in a narrow energy region in the presence of high-intensity, broad-band electronic scattering which the detector has to withstand [2,4]. In order to reduce the number non resonant photons impinging on the detector compared to the number of resonant photons, monochromators with high energy resolution and high throughput are indispensable for this technique [2,4,6,7].

Typical Mössbauer transitions studied by this technique exhibit energies in the range from 7 to 100 keV [2]. High-order reflections in silicon are used for monochromatization of x-rays in the range 7-30 keV, providing meV to sub-meV resolution [6]. Many important Mössbauer isotopes, like $^{57}$Fe, $^{151}$Eu, $^{149}$Sm, $^{119}$Sn, $^{161}$Dy, $^{201}$Hg, $^{40}$K, have transition energies in this energy range [2]. However, high-order reflections of silicon exhibit very low angular acceptance and thus low throughput at the energies higher than 30 keV [6,7]. Therefore, high energy nuclear transitions provide a challenge for constructing highly efficient high resolution monochromators. Monochromatization based on sapphire crystals in backscattering geometry had been suggested [7] and became available with meV resolution in the energy range 20-40 keV [8-10], thus providing access to studies of other interesting isotopes, like $^{125}$Te and $^{121}$Sb [8-10]. As proposed, sapphire backscattering monochromators can also be applied for meV-monochromatization in the 40-50 keV energy range [7].
At very high x-ray energies of 80-100 keV NRS studies are also possible by employing the high heat-load monochromator (HHLM) that exhibits a sufficiently small energy resolution. With this approach it was managed to observe NRS at the 89.6 keV resonance of $^{99}$Ru [11]. Although the resolution was in order of several eV in that study, NFS studies became possible due to the rather low detector efficiency and the high ratio between resonant and non-resonant events.

For the intermediate energy range of about 50 to 80 keV high-resolution-monochromators are still required to reduce the energy bandpass by about 1 to 10 orders of magnitude to prevent detector overload. This principle has been successfully applied for NFS at the 67.41 keV resonance of $^{61}$Ni [12,13] and at the 68.75 keV resonance of $^{73}$Ge [14].
Another interesting resonance is present in $^{193}$Ir at 73 keV. Currently, there is burgeoning interest in transition metal oxide (TMO) compounds containing 5d ions, such as iridium, as they afford the possibility to study the consequences of electron correlations in the strong spin-orbit coupling limit. In the weak electron correlation limit, it is well established that strong spin-orbit coupling leads to novel electronic properties, such as the topological insulator. The variety of electronic phenomena is predicted to be even richer in the presence of both significant electronic correlations and spin-orbit coupling [15]. Depending on the lattice connectivity, TMOs containing $Ir^{4+}$ ions have been predicted to realize a relativistic Mott insulator (perovskite), a quantum spin liquid with anionic excitations (honeycomb), a correlated Weyl semi-metal (pyrochlore) [15-18]. Currently, considerable experimental effort is being devoted to the study of the magnetic and electronic properties of 5d TMOs which in turn is stimulating the development of new techniques [19-22]. For example, for iridate perovskites it has recently been demonstrated that hard X-ray RIXS can be used to obtain the complete magnon dispersion curves across the entire Brillouin zone [22]. Indeed, such experiments are impossible with more conventional neutron spectroscopy given the typically small size of flux grown iridate single crystals, and the fact that iridium has a prohibitively high neutron absorption cross-section [23]. In this context, NFS as a probe of magnetism in iridates has several appealing features, including its high sensitivity and elemental specificity for determining orientation of the moments [2,4]. Due to the low absorption of 73 keV x-rays, the [193]Ir resonance itself is of particular utility as it can facilitate the use of complex sample environment such as high-field magnets and pressure cells. The latter offers the unique capability of observing the evolution of the magnetic properties of iridates through pressure driven phase transitions, which is vital to understand the relationship between the spin-orbit coupling and the electronic structure [24,25]. For instance, Donnerer et al. have reported recently the pressure-induced metallization of $Sr_3Ir_2O_7$ above 50 GPa believed to be concomitant with the quenching of the orbital moment [26].

Clearly, NFS is the only technique capable of providing definitive answers to this and other important related questions. Finally, we note that the relevant [193]Ir Mössbauer isotope exhibits a natural abundance of 62.7% [2] so that no isotopic enrichment is necessary for NFS studies.

In this paper we present a medium resolution monochromator with an energy bandpass of 300(20) meV at the [193]Ir nuclear resonance energy of 73.04(18) keV that is installed at the dynamics beamline P01 of the PETRA III synchrotron source (DESY, Hamburg). This device can be employed for hyperfine spectroscopy on Iridium and its compounds.

**Experimental aspects**

**Design and setup**

The frontend of P01 beamline is equipped with two undulators of 5 m length each and a nitrogen-cooled high-heat load monochromator (HHLM). The HHLM hosts two pairs of Si crystals, one using the (1 1 1) and another one utilizing the (3 1 1) reflection. For 73 keV x-rays the (3 1 1) reflection is used providing a photon beam with an angular divergence of 5.2(1) µrad, an energy bandwidth of 8(1) eV and a size of 2.5 x 0.75 mm$^2$ (horizontal x vertical) incident on the medium resolution monochromator (MRM).

The MRM for the $^{193}$Ir resonance is installed downstream the HHLM (Fig.1, 2). It consists of two asymmetrically cut silicon crystals. The first crystal is a dispersive element. Here, the (440) reflection is used to collimate the beam. Thus, the divergence of the beam reflected by the first crystal matches the angular acceptance of the subsequent (6 4 2) reflection in the second crystal. The main parameters of the crystals are shown in the table 1. In this configuration, the angular acceptance of the monochromator is a factor of 2.2 lower than the angular divergence of the incident beam, thus, 45% of the incident beam is accepted.

The beam spot sizes are 28.5 mm and 27 mm on the first crystal and on the second crystal respectively. The extinction length is 2.5 µm in for the first and 12.7 µm for the second crystal.

**Table 1:** Main parameters of the MRM for 73 keV x-rays

|  | First crystal | Second crystal |
|---|---|---|
| Reflection | (4 4 0) | (6 4 2) |
| Bragg angle, degree | 5.072 | 6.716 |
| Angular acceptance, µrad | 2.35 | 0.251 |
| Asymmetry parameter b | 0.11 | 2.6 |

The rather large volume of the crystal penetrated by 73 keV x-rays constitutes a challenge for the crystal quality. The two important factors which influence the quality of the crystal are the bulk purity of the silicon crystals and the thickness of the damaged layer implied during the crystal preparation process. We used silicon ingots with a resistivity ρ > 2.65 kOhm*cm.

The silicon ingots were grown by the zone melting method. Cutting of the crystals from the ingot resulted in damaged layers on the crystal surfaces of about 30 µm thickness each. Therefore, the crystals have been etched in a solution of hydrofluoric acid, acetic acid and nitric acid with a ratio of 1:2:3 volume parts. As a result 50 µm of material were removed from the surface. To achieve a planar mirror surface, the crystals have been lapped with a SiC slurry. The lapping again resulted in a damaged layer of about 20-30 µm thickness. To remove this layer, a polishing by cloth has been done afterwards. About 60 µm of material has been removed by this polishing process. Thus, we are convinced that the distorted surface layer resulting from the preparation of the crystals has been removed and has negligible impact on the performance of the crystals.

The most crucial issue in the design of the monochromator is the very small angular acceptance of the second crystal that is in the order of hundreds of nano-radians. Obviously, the mechanical setup should be able to move the crystal with the step size of this order or better. The angular positioning of the crystals is provided by two motorized stages (Fig.1) which allow step size of 24 nrad or smaller. The energy change is done by rotating of the Bragg angles of the first and second crystals with the non-equivalent step sizes given by coupling factor:

$$\frac{\delta\theta_2}{\delta\theta_1} = 3.3267 \qquad (1),$$

where $\delta\theta_1$ and $\delta\theta_2$ are the angular steps of rotation of the first and second crystal. The energy step is related to the change of the angles by the equation:

$$\delta E = -\frac{\delta\theta_2}{tan(\theta_2)+2*tan(\theta_1)} * E \qquad (2),$$

here $\theta_1$ and $\theta_2$ are the Bragg angles of the first and second crystal and $E$ is the energy of the incident x-rays.

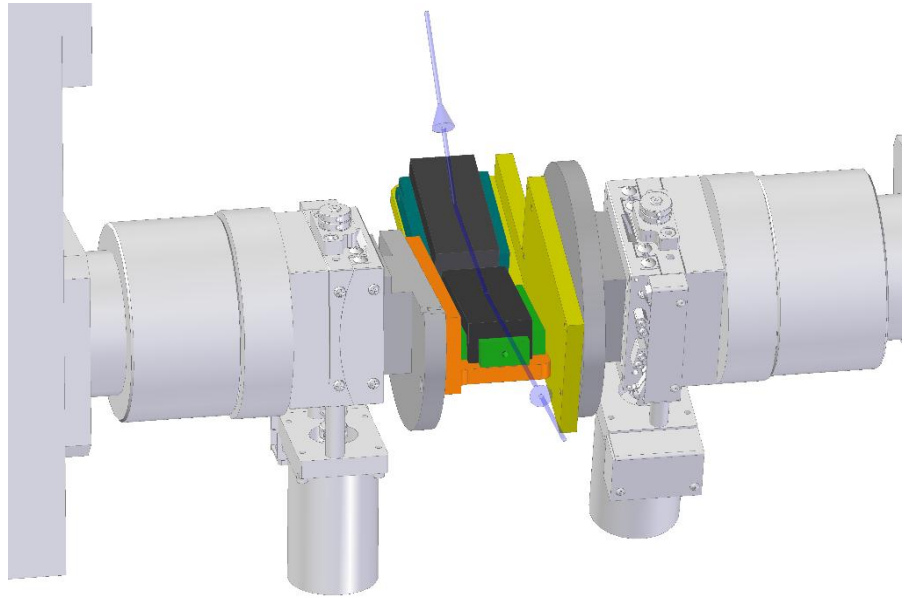

**Figure 1:** Setup of the MRM. The blue sticks with arrows indicate the beam path.

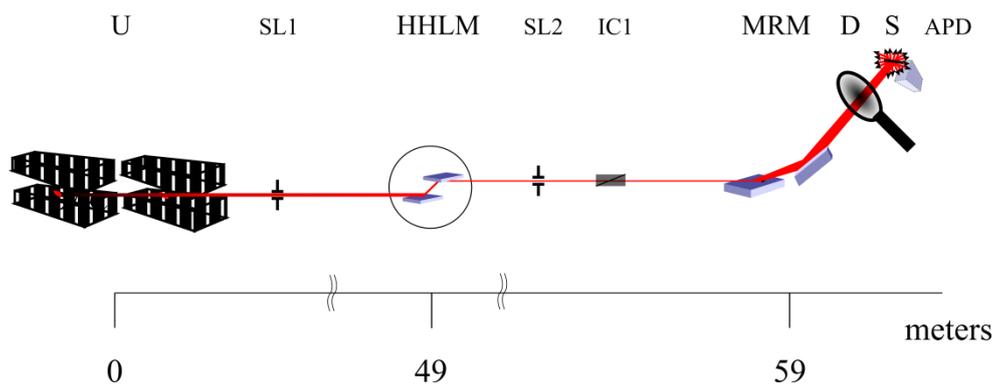

**Figure 2:** Experimental setup at the P01 beamline. U – undulators, SL1 and SL2 – slits, HHLM – high heat-load monochromator, IC1 – ionization chamber, MRM – medium-resolution monochromator, D – PIN diode detector, S – sample, APD – avalance photodiode detector.

The efficiency of the MRM is the crucial parameter for the NRS experiments, thus, it should be measured correctly. In our case the efficiency has been determined as the relation of reflected to the incident flux. The flux incident on the monochromator has been monitored by an ionization chamber. The flux of the reflected radiation was measured by a PIN diode detector and an avalanche photodiode (APD) detector has been used to verify the values obtained by the PIN diode.

The energy of the incident beam on the MRM has been pre-calibrated by Rhenium and Iridium K-edge fluorescence radiation at 71.6764(4) keV and at 76.1110(5) keV, respectively [27]. This pre-calibration was done in order to reduce the energy range to be scanned and to determine the resonance energy of $^{193}$Ir more precisely. After that the energy of the HHLM was scanned around 73 keV in order to find the $^{193}$Ir resonance. Iridium metal powder at room temperature was used to search for the resonance. The thickness of the sample was 50 μm. An APD detector with an active area of 10x10 mm$^2$ and standard NRS electronics have been used to detect the resonant photons: the signal from the APD detector has been discriminated from the detector noise by a constant fraction discriminator. The discriminated signal has been gated by the signal from bunch clock and only delayed photons between synchrotron pulses have been acquired. For more detailed information about acquisition system the reader is referred to the Refs. [2,28]. The Iridium sample was attached to the APD detector after MRM in order to observe delayed fluorescence radiation. The sample was maintained at the room temperature. Once the resonance was found, the energy of the HHLM was set to the resonance energy and the energy of MRM was scanned to refine the energy of the resonance and to measure the energy bandpass and the efficiency of the MRM.

NFS experiment desires sometimes a complex sample environment like cooling, application of strong magnetic fields or high-pressures. Notably, the beam reflected by MRM is elevated in respect to the incident beam (fig. 2). Thus, the sample can be placed before the MRM in a NFS experiment without influence on the results due to the large coherence length of radiation from Mössbauer transition (s. for instance Ref. [12]).

**Results and discussion**

We monitored the delayed fluorescence that follows the internal conversion and observed the $^{193}$Ir resonance in this scattering channel with a countrate of 6.2 Hz (Fig.3, 4). From this measurement we determined a lifetime of 8.4(2) ns for the 73 keV resonance of $^{193}$Ir which is consistent with the values reported in the literature [2,29]. The time-integrated signal was acquired in the time range between 9 and 192 ns after the synchrotron bunch. The presence of spurious bunches up to 8 ns after the main bunch in the PETRA ring during the experiment prevented an earlier start of the measurement with respect to the synchrotron bunch and did not allow us to observe the resonance in forward direction. Therefore, the energy bandpass of the monochromator has been measured by the inelastic scattering channel signal.

Most of this signal is consisting of photons produced via internal conversion with x-rays energies of 11.2, 12.8, and 13.4 keV corresponding to the L-edges of Iridium. The detection efficiency is higher at these energies than at 73 keV due to the smaller absorption length of low energy photons in the detector.

The energy and lifetime of the resonance transition have been determined to $E_0$ = 73.04(18) keV and $\tau_0$ = 8.4(2) ns, respectively. These values are in a good agreement with the literature data of $E_0$ = 73.045(5) keV and $\tau_0$ = 8.79(15) ns [2, 29].

The energy bandpass of MRM is shown in the fig. 4. Note, that the bandpass has been measured in inelastic channel, thus phonons should contribute to the measured curve. In fact, the energy bandpass function will be deconvolution of the measured curve and scattering function received from known density of phonon states (DPS) for Ir metal. Following this calculation procedure, the energy bandpass has been extracted using the DPS from the Ref. [30].

Assuming ideal silicon crystals, the dynamical theory predicts an energy bandpass of 112 meV (FWHM) and a reflectivity of 26.4% for this setup. The measured energy bandpass of 300(20) meV (FWHM) is a factor of 2.7 higher than the calculated value. The broadening can be explained by insufficient perfection of the volume of silicon crystals that is traced out by the x-ray beam. Indeed, the beam spot sizes on the crystals and extinction lengths are very high at the energy used here.

The incident flux was measured to $7.8*10^{10}$ ph/s at an electron current of 95 mA in the PETRA ring. The flux downstream the MRM was measured to $6*10^8$ ph/s. Considering these flux values and the energy bandpass functions of the HHLM and the MRM we determine a reflectivity of 9%. This value is lower than the theoretical value of 26.4% for this setup. The most reasonable explanation of the reduced reflectivity is the imperfection of the silicon crystals over the large penetration depth of the 73 keV x-ray beam.

Based on the throughput of the MRM and efficiency of the APD detectors, it is possible to estimate the countrate of NFS for Iridium metal. Iridium metal has a face-centered cubic structure and does not show any hyperfine splitting [31, 32]. From the known detector efficiency of 1.9% [33], energy bandpass of 300 meV and the flux of $6*10^8$ ph/s we expect a countrate of the NFS signal of 0.9 Hz from Iridium metal at 25 K when starting to count at 2 ns after the prompt pulse. A NFS countrate of 0.1 Hz is estimated when starting to count at 9 ns. Obviously, the two main problems of the low countrate are the late start time for counting and the efficiency of the APD detector. The former problem is related to the presence of spurious bunches in the PETRA ring. The parasitic bunches can be removed by a more efficient procedure of beam cleaning. Further, the countrate values can be improved by the use of stack of thinner APD detectors with smaller capacitance. The former assures that more resonant photons are acquired and the later makes accessible earlier starting times for counting [28].

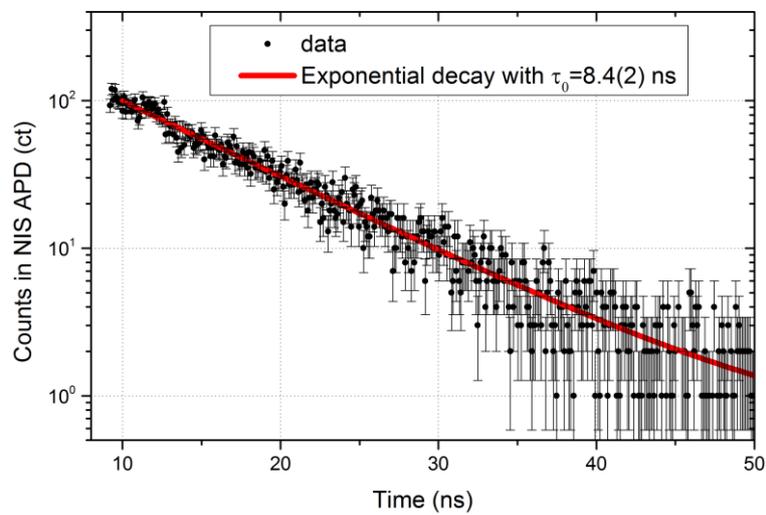

**Figure 3:** Time-spectrum of x-rays scattered in inelastic channel and exponential decay with time constant 8.4(2) ns

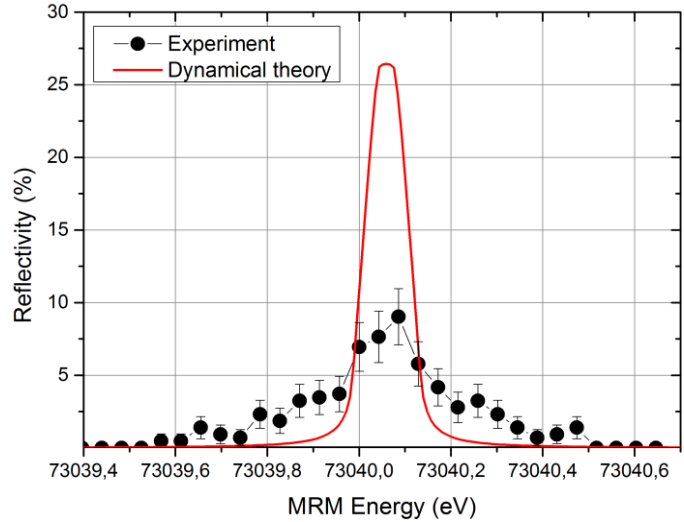

**Figure 4:** Energy bandpass of the MRM at 73.0401(3) keV measured in inelastic channel (FWHM = 300(20)meV) and predicted by dynamical theory (FWHM = 112meV)

## Conclusions

In summary, we have developed a medium-resolution monochromator for nuclear resonant spectroscopy at the 73 keV resonance of $^{193}$Ir and tested its performance. The device exhibits an energy bandpass of 300(20) meV and provides flux of $6*10^8$ ph/s at this energy. The angular acceptance of the monochromator is 2.35 µrad and angular divergence of the incident beam is 5.2(1) µrad. Thus, 45% of the incident beam is expected to be accepted. The reflectivity of 9% is smaller and the energy bandpass of 300 meV is larger than predicted by calculations. This fact can be related to the imperfection of silicon crystals over the volume penetrated by 73 keV x-rays. The imperfections are most likely resulting from distortions of crystalline structure which can be caused by impurities since the crystals did not exhibit an exceptionally high resistivity.

Using this monochromator, we have observed for the first time the nuclear resonance of $^{193}$Ir at 73.04(18) keV via excitation by synchrotron radiation. We have received a countrate of 6.2 Hz in the inelastic decay channel. The energy and lifetime, 8.4(2) ns, of the excited level are in a good agreement with the literature values. From the measured values for flux, energy bandpass and efficiency of the monochromator we predict that hyperfine spectroscopy is feasible for Iridium metal and Iridium compounds. In collaboration with the machine group of PETRA III an efficient cleaning of spurious bunches is being developed. Further, a stack of APD detectors with thickness less than 100 µm should be applied for NFS measurements [32]. Both improvements will allow for efficient time-resolved nuclear resonant scattering experiments on iridum compounds in the near future.

**Acknowledgements**   We are grateful to the Helmholtz association of German research centers, project HRJRG-402 "Sapphire ultra-optics for synchrotron radiation" for supporting this project. PETRA III synchrotron radiation source is acknowledged for provision of beam time at P01 beamline. Manfred Spiwek and the photon science crystal laboratory are greatly acknowledged for preparation of the crystals. We are thankful to the mechanical workshops of the DESY photon science department and J. Herda for the technical support.